\documentclass{article}

\usepackage{graphicx,amsmath,amssymb,chemarr,subfigure,cite,color}

\newcommand{\beq}{\begin{equation}}
\newcommand{\eeq}{\end{equation}}
\newcommand{\beqn}{\begin{eqnarray}}
\newcommand{\eeqn}{\end{eqnarray}}
\newcommand{\elabel}[1]{\label{eq:#1}}
\newcommand{\eref}[1]{Eq.\ \ref{eq:#1}}
\newcommand{\erefs}[2]{Eqs.\ \ref{eq:#1} and \ref{eq:#2}}

\newcommand{\flabel}[1]{\label{fig:#1}}
\newcommand{\fref}[1]{Fig.\ \ref{fig:#1}}

\newcommand{\avg}[1]{\left\langle#1\right\rangle}

\begin{document}

\title{Molecular clustering digitizes signaling and increases fidelity}
\author{Edward Roob III,$^{1,}$\footnote{These authors contributed equally to this work.}\,\,
Nicola Trendel,$^{2,3,}$\footnotemark[1]\,\,
Pieter Rein ten Wolde$^3$
and Andrew Mugler$^{1,3,}$\footnote{amugler@purdue.edu}\\
\\
\small{$^1$Department of Physics and Astronomy, Purdue University, West Lafayette, IN 47907, USA}\\
\small{$^2$Systems Biology Doctoral Training Centre, University of Oxford, Oxford OX1 3QU, UK}\\
\small{$^3$FOM Institute AMOLF, 1098 XG Amsterdam, The Netherlands}}
\date{}
\maketitle

\begin{abstract}
Many membrane-bound molecules in cells form small clusters.  It has been hypothesized that these clusters convert an analog extracellular signal into a digital intracellular signal and that this conversion increases signaling fidelity.  However, the mechanism by which clusters digitize a signal and the subsequent effects on fidelity remain poorly understood.  Here we demonstrate using a stochastic model of cooperative cluster formation that sufficient cooperation leads to digital signaling.  We show that despite reducing the number of output states, which decreases fidelity, digitization also reduces noise in the system, which increases fidelity.  The tradeoff between these effects leads to an optimal cluster size that agrees with experimental measurements.
\end{abstract}

\section*{Introduction}
Signaling at the cell membrane occurs in a highly organized manner.  Many membrane-bound sensory molecules are not uniformly distributed, but instead exhibit a high degree of spatial patterning.  A key example is found on the membranes of eukaryotic cells, where signaling proteins such as receptors, kinases, and GTPases form small clusters of order $5$$-$$10$ molecules \cite{Plowman2005, Tian2007, Suzuki2007a, Suzuki2007b, Ariotti2010}.  These clusters are thought to be maintained by dimer and larger complex formation \cite{Lin2014, Guldenhaupt2012, Murakoshi2004, Gurry2009}, dynamical instabilities \cite{Das2009b, Wehrens2014}, and features of the membrane environment such as cytoskeletal partitioning and lipid segregation \cite{Kusumi2005, Prior2003, Plowman2005}.  Despite the prevalence of molecular clustering, the precise role of clustering in the signaling process remains unclear.

Here we investigate the possibility that clusters improve signal transmission by discretizing a continuous input signal. Previous experimental and computational work on the Ras GTPase \cite{Tian2007, Serrano2013} and the CD59 receptor \cite{Suzuki2007a, Suzuki2007b} has suggested that clusters operate at saturation, meaning that each individual cluster produces a pulse of signaling output that is independent of the input stimulus and constant over the cluster lifetime.  This observation has led to the speculation that clusters turn an ``analog'' (smooth) signal into a ``digital'' (step-like) signal \cite{Kholodenko2010}.  However, explicit evidence of cluster-induced digitization, e.g.\ in dose-response curves, is lacking.  Previous work has also demonstrated that operating at saturation allows the total signaling output to remain a linear function of the input \cite{Tian2007, Serrano2013}.  Linearity has then been identified with high signaling fidelity, leading to the speculation that clustering increases fidelity \cite{Kholodenko2010}.  However, this definition of fidelity ignores noise in the signaling process.  Low noise is a central requirement of any high-fidelity transmitter.  Indeed, the importance of noise reduction has been recognized in the context of Ras clusters \cite{Gurry2009}, but its relation to digital signaling and transmission fidelity has yet to be elucidated.

%

Intuitively, a high-fidelity signaling system should map many input states to distinct output states with as little noise as possible (\fref{cartoon}A).  From this perspective, the expected effects of digitization on fidelity are not immediately clear.  On the one hand, digitizing an input-output curve reduces the number of output states because nearby inputs redundantly give the same output (compare the upper left and lower left panels in \fref{cartoon}A).  Therefore, just as pixelating an image reduces its recognizability, we might expect that digitizing a signal should reduce its fidelity.  On the other hand, fidelity benefits from noise reduction (compare the upper left and upper right panels in \fref{cartoon}A).  Therefore, if digitizing concomitantly reduces noise in a signaling system, we might expect that digitizing a signal should increase its fidelity.  The existence of both opposing factors makes it difficult to predict the effect of digitization on fidelity {\em a priori}.

\begin{figure}
\begin{center}
\includegraphics[width=0.9\textwidth]{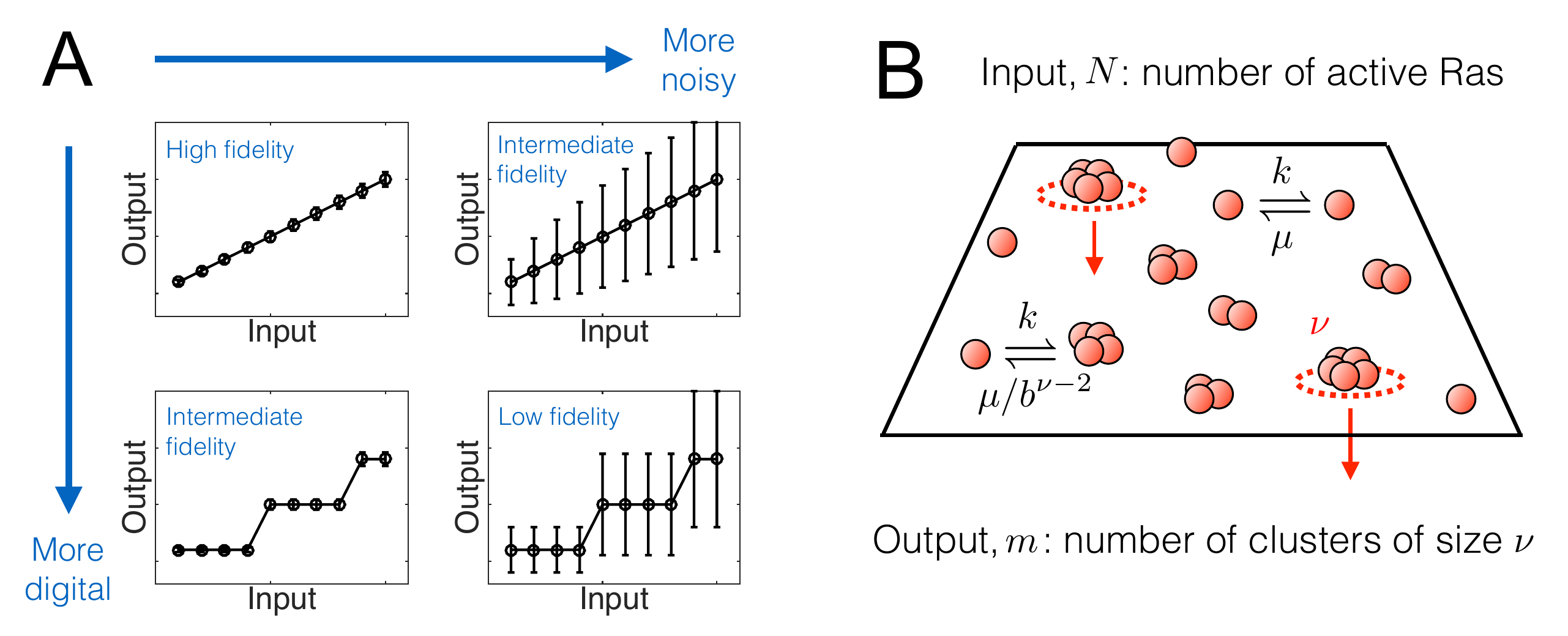}
\end{center}
\caption{Digital signaling of Ras clusters at the cell membrane. (A) Schematic illustration of the expected effects of digitization and noise on the fidelity of signal transmission. As in information theory, we define fidelity as the number of input signals that can be uniquely and reliably mapped to an output. An analog (smooth) response with low noise has the highest fidelity (upper left). Increasing noise decreases fidelity (upper right).  A digital (step-like) response also has reduced fidelity because nearby inputs lead to redundant outputs (lower left).  A digital response with high noise has the lowest fidelity (lower right). Here we explore how clustering can turn a noisy analog signal (like the upper right) into a less-noisy digital signal (like the lower left), and the associated benefits for signal fidelity. (B) Model of Ras cluster formation. $N$ active Ras molecules form clusters by sequential and cooperative monomer addition, up to a maximal size $\nu$. The $m$ clusters of size $\nu$ signal downstream.}
\flabel{cartoon}
\end{figure}

We investigate the relationship among clustering, digitization, and fidelity using theory and stochastic simulation.  
We develop a minimal model for cluster formation based on the Ras signaling system, for which several key parameters are experimentally known.  We quantify fidelity using the mutual information \cite{Shannon1948}, which captures in a natural and principled way the aforementioned benefits of increasing the state space and reducing the noise. The fidelity is then the (log of the) number of input signals that can be uniquely and reliably mapped to an output. We find that with sufficient binding cooperativity, digital signaling emerges naturally.  Associated with digital signaling is a noise reduction that initially increases the fidelity as a function of the cluster size.  Ultimately, however, when the cluster size grows large, digitization reduces the number of available output states.  The tradeoff between these two effects leads to an optimal cluster size that maximizes the signaling fidelity.  Varying the only unknown model parameter reveals that the range of optimal cluster sizes is tightly constrained and agrees well with experimentally observed ranges. This suggests, in line with our related work on spatial partitioning \cite{Mugler2013}, that protein clusters on cell membranes are tuned to maximize the transmission of signaling information.

\section*{Materials and Methods}

We consider a minimal stochastic model in which clusters $C_j$ form by the sequential addition of monomers $X$ (see \fref{cartoon}B):
\beq
\elabel{rxns}
X + X \xrightleftharpoons[\mu]{k} C_2,\qquad
X + C_2 \xrightleftharpoons[\mu/b]{k} C_3,\qquad
X + C_3 \xrightleftharpoons[\mu/b^2]{k} C_4,\qquad
\dots \qquad
X + C_{\nu-1} \xrightleftharpoons[\mu/b^{\nu-2}]{k} C_\nu.
\eeq
The first reversible reaction describes the formation and dissociation of dimers.  Subsequent reactions describe the addition of monomers one-by-one up to a maximum cluster size $\nu$.  Binding is cooperative, meaning that molecular affinity within a cluster increases with cluster size.  We assume that the association rate $k$ is diffusion-limited, and therefore cooperativity is modeled by allowing the dissociation rate to decrease with cluster size.  Specifically, dimers dissociate with rate $\mu$, and each subsequent dissociation rate is reduced by a factor $b>1$ as the cluster size increases \cite{Erdmann2009}. Thus there are two key parameters: $b$ sets the strength of cooperativity, and $\nu$ sets the maximum cluster size, independent of $b$. \eref{rxns} and similar models are well studied in the context of nucleation and growth processes, but typical treatments are deterministic (e.g.\ the Becker-D{\"o}ring equations \cite{Kashchiev2000}).  A more recent stochastic treatment has revealed interesting digitization phenomena \cite{DOrsagna2012}. Cooperativity was not considered, and importantly, the impact of signaling noise and the ensuing effects on signaling fidelity remain unexplored.

In the context of the Ras system, the monomers $X$ represent active Ras molecules.  Ras molecules dimerize on the cell membrane \cite{Lin2014, Guldenhaupt2012} with a measured dissociation constant of $\mu/k \sim 10^3$ $\mu$m$^{-2}$ \cite{Lin2014}. Ras is known to form not only dimers, but also larger clusters. Here we use the dimer data to set the basic timescale for cluster dissociation. Larger cluster formation is thought to be promoted by the presence of scaffolds \cite{Belanis2008, Shalom2008} or other proteins \cite{Murakoshi2004}, as well as the confinement of molecules within cytoskeletal or lipid domains \cite{Grecco2011}.  Stimulation of growth factor receptors increases the abundance of active Ras molecules.  Thus we take the total number $N$ of active Ras molecules as the input parameter in our model.  Our results are not sensitive to this choice: we later show that taking as our input parameter the rate of monomer activation, which is more directly a function of growth factor concentration, leads to similar results. Disruption of clustering leads to loss of downstream signaling, suggesting that the signal only propagates from clustered molecules \cite{Plowman2005, Tian2007}.  Thus we take the number $m$ of maximally sized clusters $C_\nu$ as the output of the model. Our results are also not sensitive to this choice: we later show that taking the number of clusters larger than a certain threshold as our output leads to similar results.

We simulate the reactions in \eref{rxns} in stationary state using the Gillespie algorithm \cite{Gillespie1977} (and later make analytic progress using the fact that \eref{rxns} is a closed system in equilibrium).  The two-dimensional diffusion-limited association rate is given by the measured diffusion coefficient of Ras molecules, $k \sim D \sim 1$ $\mu$m$^2$/s \cite{Lommerse2005, Lin2014, Eisenberg2011}, such that $\mu = 10^3$ s$^{-1}$.  In a given simulation we take both $\nu$ and $N$ to be fixed, although as mentioned above we later relax both of these assumptions.  Fixing $\nu$ is valid if clusters are limited in size, e.g.\ by available binding sites on a scaffold protein \cite{Belanis2008, Shalom2008}.
Fixing $N$ is valid if (i) the total number of molecules in a reaction area changes slowly, e.g.\ if molecules are confined within a domain \cite{Murase2004, Mugler2013}; and (ii) if either Ras deactivation is slow compared to clustering dynamics, or cluster formation and signaling is only weakly dependent on Ras activation state.
We perform simulations for $N$ in the range $1$$-$$N_{\max}$, where $N_{\max} = 100$ molecules (we find qualitatively similar results for different $N_{\max}$ values, with fidelity scaling as $\log N_{\max}$ in general).  Given the typical measured Ras densities of $N/A \sim 10$$-$$10^3$ molecules/$\mu$m$^2$ \cite{Lin2014, Tian2007}, this corresponds to a domain area of $A = 0.1$ $\mu$m$^2$.  We therefore set the association propensity to $\alpha \equiv k/A = 10$ s$^{-1}$.

\section*{Results}

\subsection*{Cooperativity leads to digital signaling and increased fidelity}

First we consider a single maximum cluster size $\nu$ and investigate the effects of cooperative binding.  \fref{coop}A illustrates the input-output response of our clustering model with $\nu = 7$ and for several values of the cooperativity parameter $b$.  We see that as more molecules $N$ are added to the system, more clusters $m$ form.  The number of clusters is a random variable, and therefore at a given molecule number there is a distribution of cluster numbers $p(m|N)$.  In \fref{coop}A we characterize each distribution by its mean and standard deviation, which illustrates not only the average response but also the associated noise.

\begin{figure}
\begin{center}
\includegraphics[width=0.9\textwidth]{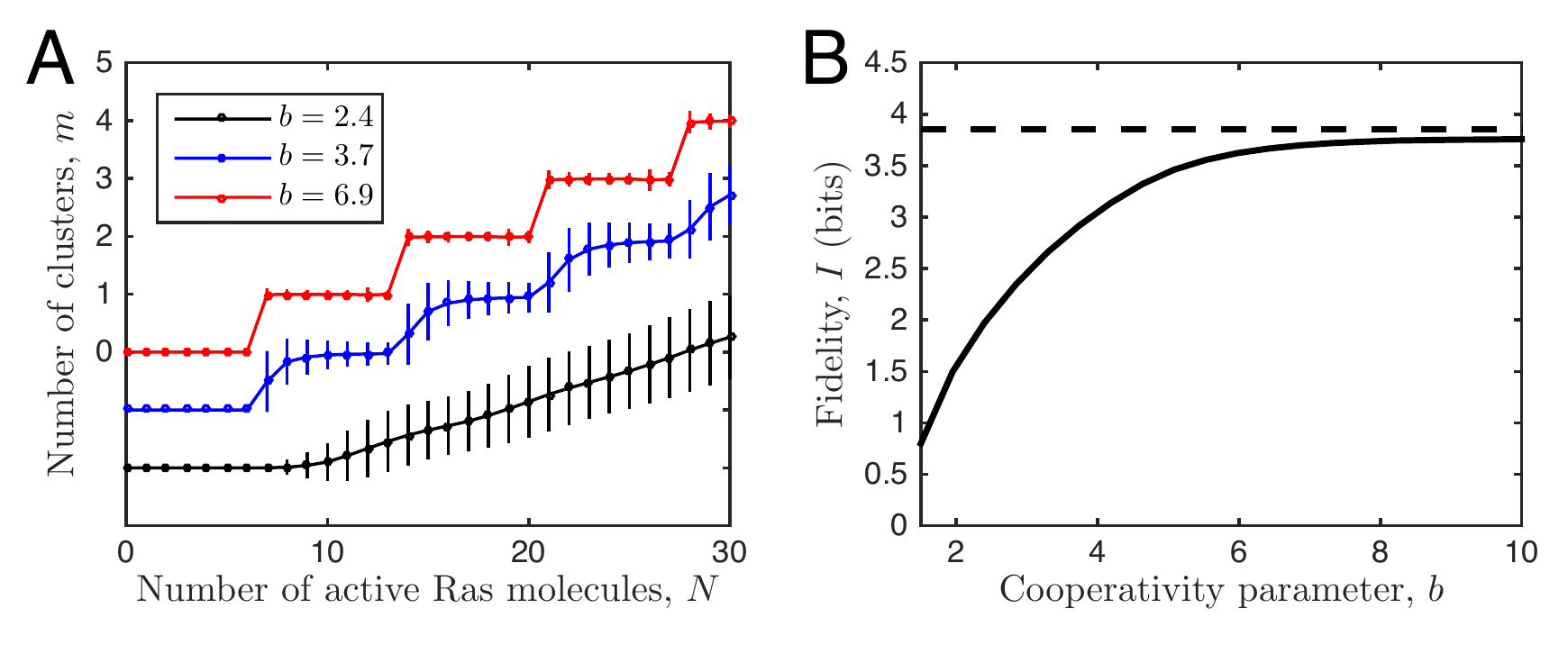}
\end{center}
\caption{Cooperativity leads to digital signaling and increased fidelity.  (A) The mean input-output response is analog (smooth) for low cooperativity (black) and digital (step-like) for high cooperativity (red).  Noise also decreases with cooperativity.  Data points and error bars show the mean and standard deviation of $p(m|N)$, respectively.  All curves start at $m=0$, but two are shifted downward for visual clarity.  (B) The fidelity increases with cooperativity.  Dashed line shows the noiseless digital limit $I = \log [(N_{\max}+1)/\nu]$.  In A and B, the maximum cluster size is fixed at $\nu = 7$, and other parameters are $N_{\max} = 100$, $\alpha = 10$ s$^{-1}$, and $\mu = 10^3$ s$^{-1}$, with simulations run for $T = 10^4$ s.}
\flabel{coop}
\end{figure}

\fref{coop}A demonstrates that at low cooperativity the response is analog (smooth), whereas at high cooperativity the response is digital (step-like).  Therefore cooperativity is required for digital signaling in our model.  The reason is that the propensity to form dimers, which is a necessary step to cluster formation in our model, is low: the forward propensity $\alpha = 10$ s$^{-1}$ is much smaller than the backward propensity $\mu = 10^3$ s$^{-1}$.  This means that among monomers and dimers alone, most molecules will be found in the monomer state.  This result is supported by observations of Ras dimerization on lipid bilayers \cite{Lin2014}.  Thus without cooperativity, even if larger clusters are allowed, the monomer state will dominate.  The effect of increasing cooperativity is to slow down dissociation from the larger clusters, and thereby push the system further into the maximally clustered state.  

Then, when the majority of molecules are found in the maximally clustered state, the number of clusters becomes highly sensitive to the total number of molecules.  Specifically, when the system contains less than $\nu$ molecules, no clusters can form, but when the system contains $\nu$ molecules exactly, one cluster forms with high probability.  This effect continues as the number of molecules increases, such that with $N$ molecules, the maximum number of clusters $z \equiv {\rm floor}(N/\nu)$ form with high probability.  The propensity for the system to form as many clusters as possible creates the digital signaling effect.

\fref{coop}A also demonstrates that in addition to digitizing the signal, increasing cooperativity reduces the noise.  This also makes sense in light of the above intuition: maximizing the propensity to cluster narrows the distribution of cluster number $m$ for a given total molecule number $N$.  In the extreme case, the system would form $z$ clusters with certainty, and the distribution $p(m|N) = \delta_{mz}$ would have zero noise. This is the key mechanism of noise reduction induced by digitization.

\fref{coop}B shows the fidelity $I$ as a function of cooperativity.  Fidelity is defined by the mutual information between the input $N$ and the output $m$ \cite{Shannon1948},
\beq
\elabel{mi}
I[N,m] = \sum_{Nm} p(m|N) p(N) \log \frac{p(m|N)}{\sum_{N'} p(m|N') p(N')}.
\eeq
We take $p(N) = 1/(N_{\max}+1)$ to be uniform, meaning that all possible values of $N$ from $0$ to $N_{\max}$ are equiprobable.  We see in \fref{coop}B that the fidelity increases with cooperativity.  This implies that for a given maximal cluster size, noise reduction outweighs the state-space reduction that is associated with digitization (\fref{coop}A).  That is, as cooperativity increases, even though fewer distinct output states can be transmitted, these states are far less noisy.  The fidelity ultimately saturates with cooperativity, since in a noiseless system with uniform input the mutual information is given by the log of the number of output states, $I = \log [(N_{\max}+1)/\nu]$.  This value is approached in the limit of high cooperativity $b$ (\fref{coop}B) and mathematically is obtained from \eref{mi} in the limit where $p(m|N) = \delta_{mz}$, and $N_{\max}+1$ is divisible by $\nu$.

\subsection*{Digital signaling results in an optimal cluster size}

We next investigate how the signaling fidelity is affected by the cluster size. \fref{opt}A shows the input-output response for several values of the cluster size $\nu$. For small $\nu$, the effect of cooperativity is insufficient to drive the system into the maximally clustered state. The input-output response is smooth and the noise is high (black curve in \fref{opt}A). The high noise corresponds to low fidelity, as shown by the blue curve in \fref{opt}B at small $\nu$. However, as $\nu$ increases, the fraction of monomers that are in the maximally clustered state increases, and the input-output relation becomes digital (blue curve in \fref{opt}A). The noise reduction associated with digital signaling leads to an increase in fidelity with $\nu$, as shown by the blue curve in \fref{opt}B at intermediate $\nu$. Finally, as $\nu$ becomes large, only a few output states are possible, since only a few clusters can be populated (red curve in \fref{opt}A). The state-space reduction leads to a decrease in fidelity with $\nu$, as shown by the blue curve in \fref{opt}B at large $\nu$. The tradeoff between noise reduction and state-space reduction leads to the optimum $\nu^*$ in \fref{opt}B at which the fidelity is maximal.

\begin{figure}
\begin{center}
\includegraphics[width=\textwidth]{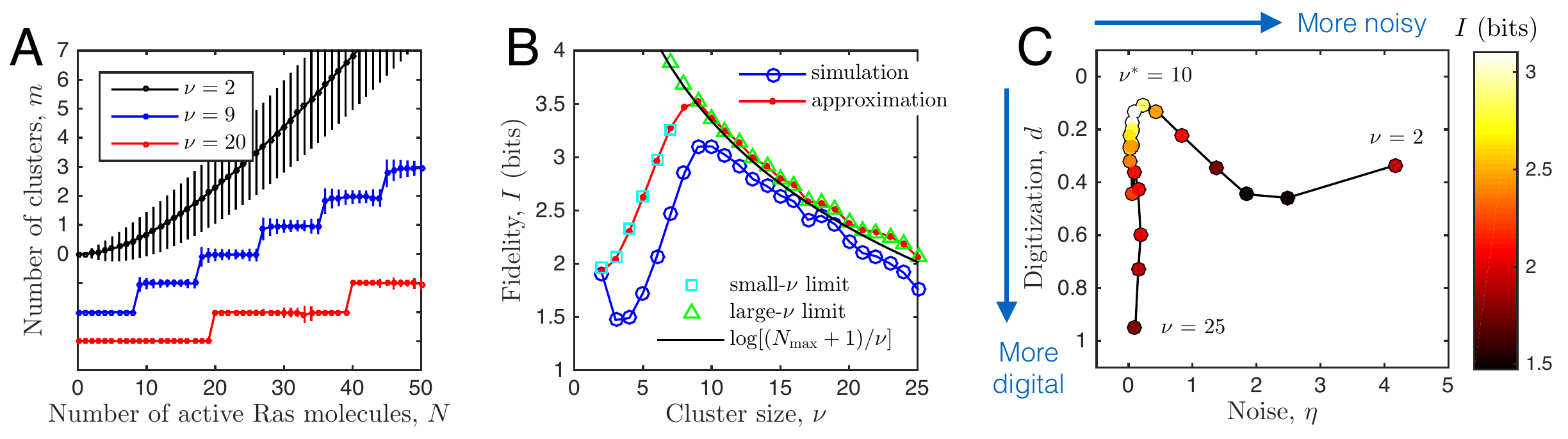}
\end{center}
\caption{Digital signaling results in an optimal cluster size.  (A) The mean input-output response is analog (smooth) for low cluster size (black) and digital (step-like) for high cluster size (red). While increasing the cluster size $\nu$ increasingly digitizes the signal, it also decreases the noise. This interplay leads to an optimal cluster size (see panel B). Data points and error bars show the mean and standard deviation of $p(m|N)$, respectively.  All curves start at $m=0$, but two are shifted downward for visual clarity. (B) Fidelity exhibits a maximum at $\nu^*$ due to the tradeoff between noise reduction and state-space reduction.  Simulation results (blue circles) are corroborated by the analytic approximation (red line, \eref{papprox}), as well as expressions in the limits of small $\nu$ (cyan squares, \erefs{mbar}{sigma2}) and large $\nu$ (green triangles, \eref{Ibumps}; black line).   (C) Increasing the cluster size moves the system through the phase space defined by digitization and noise (compare to \fref{cartoon}A).  First noise is reduced ($\eta \to 0$), then the state space is reduced, increasing digitization ($d \to 1$).  Maximal fidelity occurs at intermediate $\nu$ when noise is low, but digitization is not too high.  In all panels, parameters are $b=3$, $N_{\max} = 100$, $\alpha = 10$ s$^{-1}$, and $\mu = 10^3$ s$^{-1}$, with simulations run for $T = 10^5$ s.}
\flabel{opt}
\end{figure}

To elucidate these effects quantitatively, we derive analytic approximations to the cluster distribution $p(m|N)$ and the fidelity $I$.  First, in the limit of large $\nu$, we recognize that the system is almost entirely driven to the maximally clustered state, $p(m|N) \approx \delta_{mz}$.  Then, as above, the fidelity reduces to $I \approx \log [(N_{\max}+1)/\nu]$ when $N_{\max}+1$ is divisible by $\nu$ (\fref{opt}B, black curve).  A careful accounting of the available output states when $N_{\max}+1$ is not divisible by $\nu$ leads to the refined expression
\beq
\elabel{Ibumps}
I(\nu) = \frac{1}{N_{\max}+1}\left[(N_{\max}-\rho)\log\frac{N_{\max}+1}{\nu} + (\rho+1)\log\frac{N_{\max}+1}{\rho+1}\right],
\eeq
where $\rho$ is the remainder when $N_{\max}$ is divided by $\nu$.  \eref{Ibumps} is shown in \fref{opt}B (green triangles) and accounts for the minor bumps in the fidelity at large $\nu$ that occur due to this effect of indivisibility.

To understand the limit of small $\nu$, we seek a more general expression for the cluster distribution $p(m|N)$.  In fact, the joint distribution over the numbers of all molecule types is obtained exactly by recognizing that \eref{rxns} is a closed system in equilibrium.  The joint probability is then given by the grand canonical distribution \cite{vanKampen2007},
\beq
p(c_2,c_3,\dots,c_{\nu-1},m|N)
	= {\cal N}\frac{\lambda_2^{c_2}\lambda_3^{c_3}\dots\lambda_{\nu-1}^{c_{\nu-1}}\lambda_\nu^m}{x!c_2!c_3!\dots c_{\nu-1}!m!},
\eeq
where $c_\nu \equiv m$, the constant $\cal N$ is set by normalization, the free monomer number $x = N - \sum_{j=2}^\nu jc_j$ is set by molecule conservation, and $\lambda_j = \alpha^{j-1}/[\mu^{j-1}/b^{(j-1)(j-2)/2}]$ is set by detailed balance: it is the ratio of the product of all forward rates to the product of all backward rates governing the formation of species $C_j$.  To obtain $p(m|N)$ we would sum over all other molecule numbers, but this is difficult analytically.  Therefore we make the approximation that at sufficiently high cooperativity all molecules are either in the monomer state or the maximally clustered state, i.e.\ $c_2 = c_3 = ... = c_{\nu-1} \approx 0$.  Then,
\beq
\elabel{papprox}
p(m|N) \approx \tilde{\cal N}\frac{\lambda_\nu^m}{(N-\nu m)!m!},
\eeq
where $\tilde{\cal N}$ is a new normalization constant. Intuitively, \eref{papprox} is the stationary distribution corresponding to the effective scheme where cluster formation happens in one instantaneous step,
\beq
\elabel{eff}
\nu X \xrightleftharpoons[1]{\lambda_\nu} C_\nu.
\eeq
\fref{opt}B (red curve) shows the fidelity computed using \eref{papprox}.  We see that the approximation captures the key features seen in the simulation result, including the increase in $I$ with $\nu$ at small $\nu$, the decrease at large $\nu$, and the location of the optimum $\nu^*$.  We therefore turn to \eref{papprox} to understand the benefit of noise reduction.

To approximate the noise in \eref{papprox}, we rewrite the factorials using Stirling's approximation and expand $f(m) \equiv -\log p(m|N)$ to second order around its minimum $\bar{m}$ to obtain the Gaussian noise $\sigma^2 = 1/f''(\bar{m})$.  The result is that the mean $\bar{m}$ satisfies the equation
\beq
\elabel{mbar}
\bar{m} = \lambda_\nu(N-\nu\bar{m})^\nu,
\eeq
and $\sigma^2$ is given by
\beq
\elabel{sigma2}
\sigma^2 = \frac{\bar{m}(N-\nu\bar{m})}{N+\nu(\nu-1)\bar{m}}.
\eeq
\eref{mbar} is expected because it is the steady state of the macroscopic rate equation corresponding to \eref{eff}.  \erefs{mbar}{sigma2} are validated in \fref{opt}B by confirming that at small $\nu$ the fidelity computed numerically from the Gaussian approximation (cyan squares) agrees with that calculated analytically from \eref{papprox} (red curve).

Because \eref{mbar} is a polynomial equation of arbitrary degree, it is difficult to find a general expression for $\bar{m}$.
Nonetheless, for small $\nu$ when the number of clusters is far from its maximum value $\bar{m} \sim N/\nu$, we can make the approximation $\epsilon \equiv \nu\bar{m}/N \ll 1$.  Expanding \erefs{mbar}{sigma2} to first order in $\epsilon$ gives $\bar{m} = \lambda_\nu N^\nu(1-\nu\epsilon)$ and $\sigma^2 = \lambda_\nu N^\nu(1-2\nu\epsilon)$, and therefore a signal-to-noise ratio of
\beq
\elabel{snr}
{\rm SNR} = \frac{\bar{m}^2}{\sigma^2} = \lambda_\nu N^\nu + \mathcal{O}(\epsilon^2).
\eeq
Recalling that $\lambda_\nu = [\alpha b^{(\nu-2)/2}/\mu]^{\nu-1}$, we obtain ${\rm SNR} = [\alpha b^{(\nu-2)/2}/\mu]^{\nu-1} N^\nu$. This expression is an increasing function of $\nu$ when $\nu > \nu_0 \equiv \log(\mu b^{3/2}/\alpha N)/\log b$, which is easily verified by differentiating the log.  For example, setting $N = 1$$-$$100$ and $b=3$ produces the range of small values $\nu_0 = 1.5$$-$$5.7$.  Thus we see that after a brief decrease, the SNR increases with $\nu$.  This behavior is consistent with the dependence of the fidelity on $\nu$ seen in \fref{opt}B for $\nu < \nu^*$.  Indeed, this behavior is precisely the effect of noise reduction: except at the smallest $\nu$ values, increasing $\nu$ allows for higher cooperativity, thereby reducing the noise and causing the SNR (and in turn the fidelity) to increase with cluster size.

Finally, we demonstrate in \fref{opt}C how varying the cluster size moves the system through the phase space suggested by \fref{cartoon}A.  We compute the average noise in the response $\eta = \avg{\sigma^2}$ and quantify the degree of digitization as $d = 4\avg{|\bar{m}-a|}/\bar{m}_{\max}$, where $a \equiv \bar{m}_{\max}N/N_{\max}$. In these expressions, averages and maxima are over $N$, and as above $\bar{m}$ and $\sigma^2$ are the $N$-dependent mean and variance of $p(m|N)$. The digitization $d$ is the average deviation from the straight line $\bar{m} = a$ from $(0,0)$ to $(\bar{m}_{\max},N_{\max})$, with the normalization $4/\bar{m}_{\max}$ ensuring that $d$ ranges from $0$ to $1$. The value $d=0$ corresponds to a mean response $\bar{m} = a$ that is completely analog and linear, while the value $d=1$ corresponds to a mean response $\bar{m} = \bar{m}_{\max}\theta(N > N_{\max}/2)$ that is a maximally digital two-state switch.

\fref{opt}C demonstrates that as the cluster size increases, first the noise is reduced ($\eta \to 0$), then the state space is reduced ($d \to 1$).  Optimal fidelity is achieved for an intermediate cluster size ($\nu^* = 10$), when the noise is low, but the digitization is not so severe that too few states are available.  Such a ``fine'' degree of digitization, wherein each cluster signals digitally but there are many clusters (see the blue curve in \fref{opt}A), has been described as a ``digital-analog'' system, such that clusters act collectively as an ``analog-to-digital-to-analog'' (ADA) converter of an external analog signal \cite{Tian2007, Kholodenko2010, Zhou2014}.  Our results demonstrate explicitly and quantitatively that this ADA regime is optimal for signaling fidelity.

\subsection*{Optimum is robust to model assumptions}

We now relax both model assumptions, namely that $\nu$ and $N$ are fixed. Allowing $\nu$ to vary means that signaling is no longer restricted to only the clusters of maximal size $C_\nu$. Instead, we allow clusters of many sizes $C_j$ to signal. Specifically, we define a threshold $j_{\min}$ such that all clusters with $j \ge j_{\min}$ can signal, and we then vary $j_{\min}$. We continue to use $m$ to denote the number of clusters that signal downstream, and we compute the fidelity $I[N,m]$ as before. \fref{robust}A shows that the presence of an optimum $\nu^*$ is robust to the choice of $j_{\min}$. In fact, the optimum persists even when $j_{\min}$ is as low as $0.3\nu$, for which clusters of almost any size can signal downstream.

\begin{figure}
\begin{center}
\includegraphics[width=0.9\textwidth]{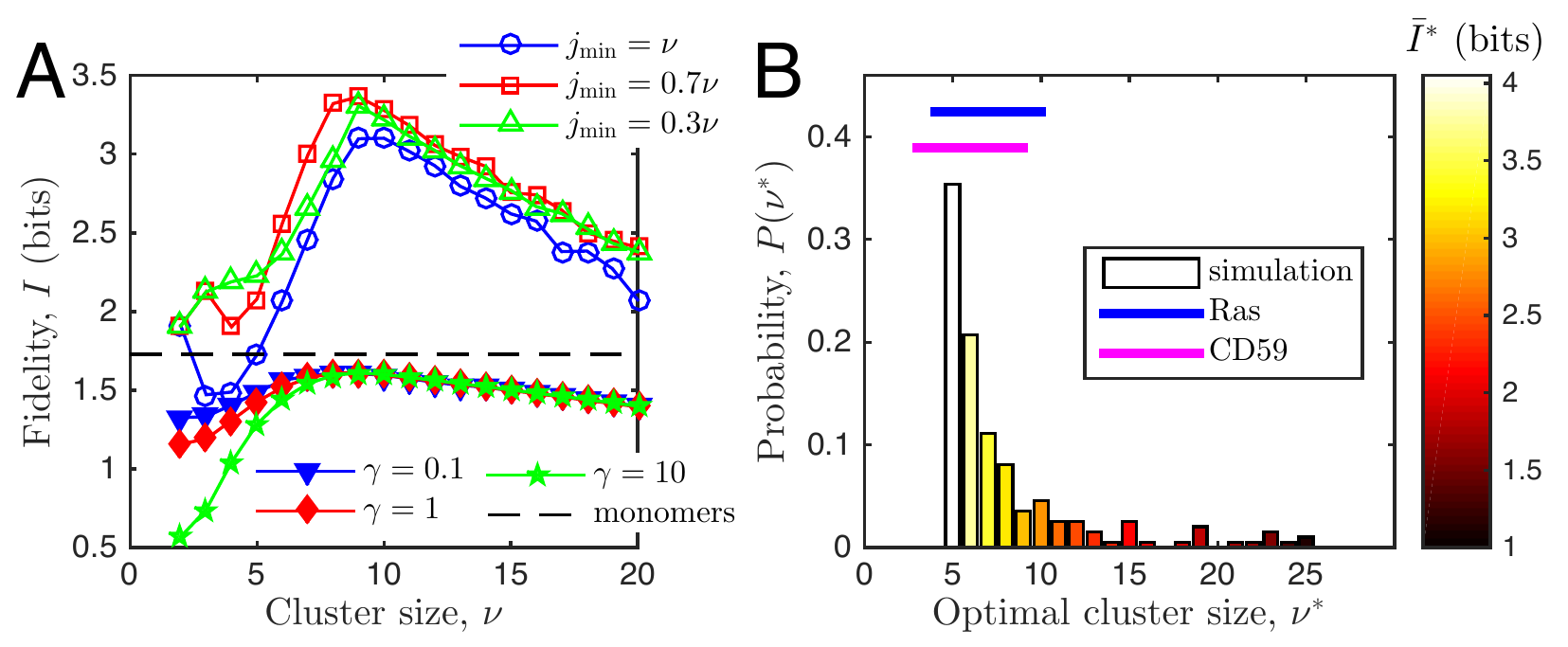}
\end{center}
\caption{Optimal cluster size is robust to model assumptions and agrees with experiments.  (A) Optimum persists both when (i) we vary the minimal cluster size $j_{\min}$ that can propagate the signal downstream (open symbols) and (ii) we allow the total number of active Ras molecules to fluctuate, with deactivation rate $\gamma$, adding input noise to the system (filled symbols). Black dashed line shows fidelity when only monomers are present (see Discussion). (B) Distribution of optimal cluster size $\nu^*$ for $100$ cooperativity values sampled from $b = 1.25$$-$$100$ (uniformly in log space).  The most frequent $\nu^*$ values also have the highest mean fidelity (colors) and are contained within the range $5 \le \nu^* \lesssim 10$, in good agreement with the observed ranges for Ras ($6$$-$$8$ \cite{Plowman2005} and $4$$-$$10$ molecules \cite{Janosi2012}) and CD59 clusters ($3$$-$$9$ molecules \cite{Suzuki2007a, Suzuki2007b}).  Parameters are: in A, $\beta = 3$ (filled symbols), $b=3$, and $T = 10^4$ s; in B, $T = 10$ s (sufficient to identify the optima); and in both, $N_{\max} = 100$, $\alpha = 10$ s$^{-1}$, and $\mu = 10^3$ s$^{-1}$.}
\flabel{robust}
\end{figure}

Allowing $N$ to vary corresponds to accounting for the upstream process by which Ras molecules are activated and deactivated.
For simplicity, we incorporate upstream activation using the approximate model in \eref{eff}. We allow Ras molecules to be activated from a large pool of inactive molecules at a constant rate, in bursts of size $\beta$, and deactivated at a constant rate $\gamma$. Deactivation can occur within a cluster, at which point the cluster dissociates into the remaining $\nu-1$ active monomers. Together with \eref{eff}, we have
\beq
\elabel{inputnoise}
\emptyset \xrightarrow{N\gamma/\beta} \beta X,\qquad
X \xrightarrow{\gamma} \emptyset,\qquad
C_\nu \xrightarrow{\nu\gamma} (\nu-1)X.
\eeq
The first two reactions describe the gain and loss of active Ras molecules $X$, respectively. The activation rate $N\gamma/\beta$ is defined such that the mean number of $X$ molecules is $N$. Therefore, $N$ is now the mean of a fluctuating number of molecules, instead of the fixed total number of molecules. The third reaction describes the dissociation of a cluster when any one of the $\nu$ monomers deactivates with rate $\gamma$. As before we compute the fidelity $I[N,m]$.

\fref{robust}A shows that the optimum persists in this model with upstream activation dynamics. The overall value of the fidelity is reduced in this model, which makes sense since the upstream activation process introduces additional input fluctuations into the system. We also see that fidelity decreases as $\gamma$ increases, because larger $\gamma$ promotes cluster break-up via the third reaction in \eref{inputnoise}. Fewer clusters means a narrower output range, and thus a decreased fidelity. While increasing $\gamma$ decreases fidelity because of cluster break-up, we also find that larger $\gamma$ makes clustering more beneficial than signaling via the monomers themselves, as detailed further in the Discussion. Altogether, \fref{robust}A demonstrates that the presence of an optimal cluster size $\nu^*$, as well as its value, is robust to the model assumptions.

\subsection*{Optimal cluster size agrees with experiments}

What sets the value of the optimal cluster size $\nu^*$?  The only unknown parameter in the model of \eref{rxns} is the cooperativity $b$, and indeed $\nu^*$ varies with $b$.  However, even when $b$ is varied over
two orders of magnitude,
the majority of high-fidelity $\nu^*$ values remain within the range $5 \leq \nu^* \lesssim 10$ (\fref{robust}B).  The lower bound occurs because even for very large $b$, a minimum number of binding events is required for cooperativity to take effect and drive the system to the clustered state.  The upper bound occurs because although very small $b$ leads to larger values of $\nu^*$, these solutions then suffer from low fidelity due to the  unavoidable state-space reduction at large $\nu$ (\fref{opt}B).  Together these effects keep $\nu^*$ tightly constrained within the range $5 \leq \nu^* \lesssim 10$.  Interestingly, this range agrees remarkably well with cluster sizes observed in experiments, both for Ras and the CD59 protein (\fref{robust}B).

It is worth emphasizing that \fref{robust}B contains no free parameters.  All parameters that are not optimized or varied are taken from measurements of Ras density, diffusion, and dissociation.  Therefore, the agreement of the model predictions with the experimental cluster size data in \fref{robust}B presents strong evidence that clustered sensory molecules are tuned to maximize fidelity via digital signaling.

\section*{Discussion}

We have seen that cooperative molecular clustering leads to digital signaling and increased signaling fidelity.  The increase in fidelity is surprising, because digitization reduces the number of available output states, which makes many input values redundant.  Alone, this effect would reduce fidelity.  However, this effect is compensated by a noise reduction that accompanies digital signaling.  Noise reduction increases fidelity.  The tradeoff between state-space reduction and noise reduction results in an optimal cluster size that maximizes signaling fidelity.  The range of optimal cluster sizes predicted by the model with no free parameters agrees strikingly well with data on the Ras and CD59 systems.


In our model, the number of clusters $m$ increases roughly linearly with the total number of molecules $N$, both for analog and digital signals (\fref{coop}A and \fref{opt}A).  It has been argued that an equilibrium model, such as \eref{eff}, would not yield a linear relationship because mass-action kinetics imply a sharp polynomial increase of $m$ with the free monomer number $x$ \cite{Tian2007, Gurry2009}.  Indeed, this polynomial dependence appears directly in \eref{mbar}.  However, because the total molecule number is fixed, $x + \nu m = N$, even though $m$ increases sharply with $x$, it does not necessarily increase sharply with $N$.  Instead, we find that $m$ increases roughly linearly with $N$. Therefore, our results imply that if the total molecule number is fixed or varies sufficiently slowly, an equilibrium model can produce a linear input-output relationship, which has been argued to be
necessary to explain clustering data in the Ras system \cite{Plowman2005, Tian2007}.

Our model predicts that due to cooperativity, the typical cluster lifetime is longer than the typical dimer lifetime.  We estimated the dimer lifetime from Ras dissociation and density data \cite{Lin2014} to be $\mu^{-1} \sim 1$ ms.  On the other hand, the cluster lifetime is set by the rate-limiting step for cluster dissociation, $\tau \sim (\mu/b^{\nu-2})^{-1}$ (\eref{rxns}).  This time is longer than $1$ ms: for example, with typical values $\nu = \nu^* \sim 7$ and $b \sim 3$ we would predict $\tau \sim 0.3$ s.  Indeed, cluster lifetimes inferred from transient immobility data are also on the order of tenths of seconds, not a millisecond: for Ras $\tau \approx 0.1$$-$$1$ s \cite{Murakoshi2004, Zhou2014} and for CD59 $\tau \approx 0.57$ s \cite{Suzuki2007a, Suzuki2007b}. This agreement with experiments supports the model presented here, and it also raises the interesting question of what are the benefits and limitations of longer-lived clusters.

We have found that an optimal cluster size in the range of about $5$ to $10$ molecules allows for higher fidelity than any other cluster size. Does it allow for higher fidelity than a strategy of not clustering at all, and instead signaling using monomers only? To investigate the scenario in which signaling proceeds via monomers (instead of larger clusters), we consider a system consisting of only the reactions in \eref{inputnoise}, thus without those describing cluster formation. The output $m$ in this scenario is simply the active monomer number, and the fidelity is computed from the stationary distribution $p(m|N)$. In \fref{robust}A, for example, the fidelity is $I \approx 1.73$ bits (black dashed line). This is higher than the fidelity with clustering (filled symbols), meaning that clustering is not the optimal strategy. In general, we find that clustering is only the optimal strategy at large $\beta$, $\gamma$, and $\lambda_\nu$ in \erefs{eff}{inputnoise}. Large $\beta$ is expected because then monomer activation is bursty and thus noisy. Large $\gamma$ is also expected because then the monomer activation dynamics are faster than the cluster formation dynamics, and clustering integrates over the upstream noise. The benefit of integration persists despite the fact that large $\gamma$ promotes cluster break-up as described in the previous section. Finally, large $\lambda_\nu$ keeps the mean cluster number high, which is necessary for a large output range and thus high fidelity. Outside of this regime, clustering may have additional advantages over using only monomers, such as introducing a delay from a millisecond to tenths of a second as described above, which could prevent premature propagation of a spurious input signal. Given that clustering occurs, our study shows that an optimal cluster size exists, is robust to modeling assumptions, is tightly constrained, and agrees with experimental estimates.

Our results relate to recent findings on the accuracy of sensing a single, constant input concentration using cooperative cluster formation. In equilibrium systems, such as the system studied here (\eref{rxns}), it has been shown that cooperativity increases sensing accuracy \cite{Govern2014}. Indeed, we also find here that accuracy is maximized in the limit of maximal cooperativity (\fref{coop}B). It was also found that in equilibrium systems, accuracy is maximized in the limit of maximal cluster size \cite{Govern2014}, which would correspond to $\nu = N_{\max}$ here. However, we do not find $\nu = N_{\max}$ here. Instead we find a finite optimal $\nu^*$. The reason is that we do not maximize the accuracy of sensing a single input, but rather a range of inputs. While a large cluster size allows a system to measure one input very reliably, it also reduces the number of inputs that can be measured. This tradeoff leads to the optimal cluster size $\nu^*$. Finally, in contrast to purely equilibrium systems, it has been shown in the context of an Ising-like model of receptor activation in the bacterial chemotaxis system that in non-equilibrium systems that time-integrate the receptor dynamics \cite{Skoge2011, Skoge2013}, cooperativity does {\it not} improve sensing accuracy. The reason is that cooperativity not only increases the gain, but also slows down the receptor dynamics. The slowing down then actually decreases the time-averaged accuracy \cite{Skoge2011, Skoge2013}, unlike in equilibrium systems, which do not time-integrate the receptor dynamics but rather rely on an instantaneous readout of the receptor \cite{Govern2014}. As mentioned above, we also find here that clustering slows down the dynamics, and for the same reason this may hamper time-integrated sensing accuracy downstream. We leave the question of how the downstream system integrates the Ras clustering dynamics as an interesting topic for future work.

The mechanism we investigate likely integrates with other features of the Ras system that contribute to high-fidelity signaling.  For example, it is been shown that enzyme-mediated positive feedback leads to digital, two-state activation of Ras \cite{Das2009a}, which could very well enhance the subsequent cluster-induced digitization elucidated here.  Moreover, confinement alone, even without cooperative binding, has been shown to increase the fidelity of Ras signaling \cite{Mugler2013}, and clustering could therefore further enhance this effect.  Clustering also enhances downstream signal propagation in the Ras-mediated mitogen-activated protein kinase cascade \cite{Mugler2012}.
It will be interesting to explore how these mechanisms are integrated, as well as the extent to which cluster-driven digitization occurs across other domains of biology.


\section*{Acknowledgements}
We thank Nils Becker and Aleksandra Walczak for useful comments on the manuscript.

\bibliographystyle{unsrt}
\bibliography{refs}

\end{document}